# Scalability Analysis of Distributed Kolmogorov-Arnold Network Training on High-Performance Computing Systems

**Guangneng Chen**[a,*], **David García-Selfa**[b], **Pablo Quesada Barriuso**[a]

[a] *University of Santiago de Compostela, 15782 Santiago de Compostela, Spain*

[b] *Galicia Supercomputing Center (CESGA), 15705 Santiago de Compostela, Spain*

* Corresponding author. E-mail address: guangneng.chen@rai.usc.es

## Abstract

Kolmogorov-Arnold Networks (KANs) replace the fixed activation functions and linear weights of Multi-Layer Perceptrons (MLPs) with learnable univariate functions on network edges, offering improved interpretability and, in some settings, competitive parameter efficiency. While the approximation properties of KANs have received considerable attention, their behavior under distributed, multi-GPU training has not been systematically characterized. This paper presents an empirical scalability study of data-parallel KAN training on multi-node, multi-GPU high-performance computing (HPC) infrastructure, evaluated along four dimensions: strong scaling, weak scaling, communication overhead, and model-size scaling.

Experiments were conducted on the FinisTerrae III supercomputer using up to 8 NVIDIA A100 GPUs across 4 nodes with PyTorch Distributed Data Parallel (DDP). KAN training reaches 74.7% parallel efficiency at 8 GPUs with a 5.97× speedup, consistent with conventional deep learning workloads rather than a KAN-specific advantage or penalty. Weak scaling shows an initial single-to-multi-GPU throughput drop followed by strong stability, with a coefficient of variation under 0.3% among distributed configurations. Communication overhead follows a non-monotonic pattern across topologies (1.3%–6.1%), driven primarily by All-Reduce algorithm selection and inter-node latency rather than KAN's edge-wise gradient structure. The parameter-to-memory ratio improves with model size (74K→118K parameters/GB) even as per-parameter throughput and training time scale unfavorably, growing 7.08× for an 8× increase in width.

These results indicate that operator-level and data-parallel optimizations for KAN are complementary rather than competing. We provide deployment guidelines for GPU topology and model-size selection, and discuss the limitations of a synthetic-regression, single-cluster evaluation, outlining the real-workload validation needed to generalize these findings.

## 1. Introduction

The field of deep learning has witnessed remarkable progress in recent years, with neural networks achieving strong performance across domains including computer vision, natural language processing, and scientific computing. Despite its widespread adoption, the traditional Multi-Layer Perceptron (MLP) architecture faces recognized limitations in interpretability and parameter efficiency, since its learnable weights are linear and its activation functions are fixed.

Kolmogorov-Arnold Networks (KANs) [1] depart from this design by placing learnable, spline-parameterized univariate functions on network edges, while nodes perform simple summation. This construction is motivated by the Kolmogorov-Arnold representation theorem [2, 3], which shows that any continuous multivariate function can be expressed as a finite composition of continuous univariate functions. Formally, for any continuous function $f$ on $[0,1]^n$, there exist continuous univariate functions $\Phi^{q,p}$ and $\Psi^q$ such that

$$f(x_1, \ldots, x_n) = \sum_{q=0}^{2n} \Psi_q \left( \sum_{p=1}^{n} \Phi_{q,p}(x_p) \right)$$

This theorem provides the theoretical foundation for constructing networks whose learnable components are placed on edges as univariate functions rather than as node-level linear weights combined with fixed pointwise nonlinearities, as in standard MLPs. Fig. 1 contrasts the two designs directly: an MLP composes fixed nonlinearities σ applied at nodes with learnable linear weights W on edges, MLP(x) = (W3∘σ2∘W2∘σ1∘W1)(x), whereas a KAN composes learnable univariate functions Φ on edges with simple summation at nodes, KAN(x) = (Φ3∘Φ2∘Φ1)(x).

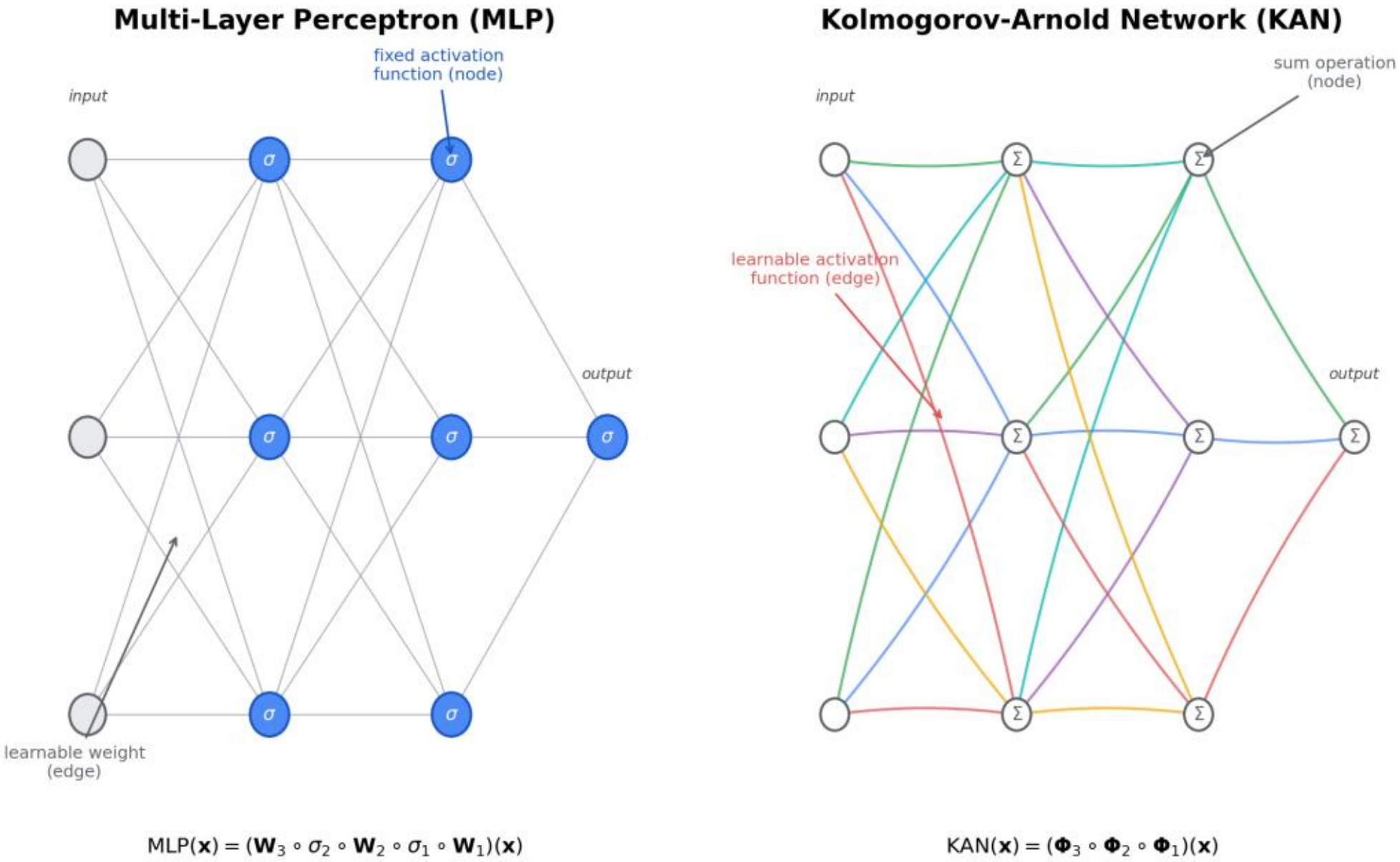

**Fig. 1. Architectural comparison between MLP and KAN: fixed node-wise activation with learnable edge weights (left) versus learnable edge-wise activation with node-wise summation (right).**

KANs have attracted interest for their interpretability and parameter efficiency relative to MLPs [1, 4], but the computational cost of evaluating and differentiating B-spline bases at every edge raises a natural question that has so far been addressed from two different angles in the literature.

One line of work optimizes the intra-operator computation of KAN itself. MatrixKAN [5] reformulates the recursive Cox-de Boor B-spline evaluation as batched matrix operations, reporting up to a ≈40× speedup over the reference PyKAN implementation on a single device for high-degree B-splines. The reference implementation's per-step cost scales linearly with spline degree, while MatrixKAN's does not, so the realized speedup depends on the spline order used. Polar and Poluektov [6] address a different training regime for KAN – the Newton-Kaczmarz (NK) fitting method. They propose pre-training, disjoint-dataset partitioning with model merging, and an FPGA-targeted parallelization to work around the sequential dependency between successive NK parameter updates. Both target the computation performed within a single KAN training run on a single device or accelerator, rather than the replication of a KAN model across multiple GPUs and nodes.

A second, complementary question is how a KAN model behaves once it is replicated across multiple GPUs and nodes using standard data-parallel training – the setting in which the vast majority of production-scale neural network training takes place. This question has an established methodology in the HPC and distributed-systems literature (strong/weak scaling, communication overhead, resource utilization), but it has not been applied specifically to KAN. KAN's learnable activations are stored and updated per-edge rather than per-node. It is therefore not obvious a priori whether this architectural difference translates into distinguishable communication or memory-scaling behavior relative to MLP-style networks. The dominant costs could instead simply be those of any DDP-trained model with a comparable parameter count.

This paper addresses that question empirically. We conduct a systematic scalability study of DDP-based KAN training on the FinisTerrae III supercomputer, evaluating strong scaling, weak scaling, communication overhead, and model-size scaling across configurations from a single GPU to 8 GPUs spanning 4 nodes. Rather than assuming that KAN's spline-based computation must produce architecture-specific distributed-training behavior, we test this assumption directly and report where the observed scaling patterns are consistent with general DDP behavior and where they are not.

The contributions of this work are: (1) an empirical characterization of strong- and weak-scaling behavior for KAN under DDP, reaching 74.7% parallel efficiency and 5.97× speedup at 8 GPUs; (2) a topology-resolved analysis of communication overhead that identifies All-Reduce algorithm selection and node placement, rather than KAN's architecture, as the primary source of the observed non-monotonic overhead pattern; (3) a model-size scaling study relating parameter count to throughput, training time, and the parameter-to-memory ratio, showing that larger KAN models attain a more favorable ratio of parameters to memory usage even as training time grows faster than a fixed per-sample cost would predict; and (4) deployment guidelines for GPU topology and model-size selection, explicitly positioned as complementary to operator-level optimizations such as MatrixKAN rather than as a substitute for them.

The remainder of this paper is organized as follows. Section 2 reviews related work in distributed deep learning and KAN parallelization, and positions this study relative to it. Section 3 describes the experimental methodology, platform, and configurations. Section 4 presents results on strong scaling, weak scaling, communication overhead, and model scaling (Sections 4.1–4.4, respectively). Section 5 synthesizes these results and discusses optimization strategies, deployment guidance, the relationship of this study to operator-level KAN acceleration, and limitations. Section 6 concludes.

# 2. Related Work

## 2.1. Data-Parallel Distributed Training

Data parallelism is the most widely used strategy for distributed neural network training. Goyal et al. [7] showed that large-batch synchronous SGD can scale near-linearly with proper learning-rate warmup, enabling ImageNet training in one hour across hundreds of workers — a synchronous, data-parallel regime directly relevant to the DDP setting studied here. Alternative directions relax this synchrony or communication cost: Chen et al. [8] proposed synchronous SGD with backup workers to mitigate the straggler problem without incurring asynchronous noise, and Alistarh et al. [9] introduced gradient quantization (QSGD) to reduce communication volume by up to 32×; this paper instead evaluates the standard synchronous, non-quantized DDP setting. Li et al. [10] documented the engineering principles behind PyTorch's DDP implementation, which we use in this study, and Patarasuk and Yuan [11] established bandwidth-optimal All-Reduce algorithms whose design principles are reflected in modern collective communication libraries such as NCCL, whose topology-dependent behavior we revisit empirically in Section 4.3. For very large models, model-parallel techniques such as tensor parallelism [12] and pipeline parallelism, together with memory-redundancy elimination [13], extend training beyond what data parallelism alone can support. Standardized benchmarking efforts such as MLPerf [14] and DeepBench [15] motivate the strong-scaling, weak-scaling, and communication-overhead evaluation methodology that this paper applies to KAN.

## 2.2. Parallelization Approaches for KAN

Two recent lines of work address parallelization specifically for KAN, and it is important to state clearly how the present study relates to them.

Operator-level and algorithm-level parallelization. MatrixKAN [5] reformulates the recursive Cox-de Boor B-spline evaluation as batched matrix multiplications, removing its sequential dependency between spline segments. This reformulation reaches up to ≈40× forward/backward speedup on a single device for high-degree splines. Polar and Poluektov [6] instead target KAN trained via the Newton-Kaczmarz (NK) method, where each parameter update depends sequentially on the previous one. They propose pre-training, disjoint-dataset partitioning with model merging, and an FPGA-based parallelization of the NK update, reporting substantial training-time acceleration over sequential execution. Both works accelerate the computation performed within a single KAN training run on a single device, rather than the data-parallel replication of a KAN model across a distributed multi-GPU, multi-node cluster – the question this paper addresses.

System-level distributed training. The present study addresses a different, complementary layer of the problem: given an already-instantiated KAN model (using the reference recursive spline evaluation, without operator-level acceleration), how does training scale when that model is replicated across multiple GPUs and nodes using standard data-parallel DDP? This includes communication overhead from gradient All-Reduce, memory scaling with model size, and throughput scaling with GPU count – questions that are orthogonal to how efficiently a single spline evaluation is computed.

These two directions are not competing but additive. An operator-level speedup such as MatrixKAN reduces the per-GPU computation time Tcompute in each training step; the present study characterizes how the communication time Tcomm that must be added to it scales with GPU count and topology. Because parallel efficiency in a DDP setting is approximately Ep ≈ Tcompute/(Tcompute + Tcomm), a faster per-GPU kernel (lower Tcompute) would, all else equal, make the communication overhead documented in Section 4.3 of this paper a larger relative fraction of step time, not a smaller one – meaning that adopting operator-level optimizations such as MatrixKAN inside a DDP pipeline would make communication-aware topology selection more rather than less important. We return to this interaction in Section 5.4.

## 2.3. Function-Approximation Architectures

KANs belong to a broader family of neural architectures built on structured basis-function approximation rather than affine transformations followed by pointwise nonlinearities, including networks built on adaptive or learnable activation functions more generally [16, 17]. The mathematical theory of splines [18] underlies KAN's edge functions and is directly relevant to the memory and computation patterns analyzed in Sections 4.3 and 4.4.

## 2.4. Positioning of This Work

Building on this literature, the present study makes an empirical, systems-oriented contribution: a scalability characterization of data-parallel KAN training under DDP on real multi-node HPC hardware, using the standard strong-scaling, weak-scaling, and communication-overhead methodology from the distributed-training literature [14], applied here, to the best of our knowledge, for the first time to a KAN model. We do not modify or accelerate KAN's internal spline computation – that problem is addressed by MatrixKAN [5] and related work [6] – and our results should be read as characterizing the reference (unoptimized) KAN operator under distributed replication. Section 5.4 discusses how the two lines of work could be combined in future systems.

## 3. Methodology

This section describes the experimental design, performance metrics, platform, and configurations used to evaluate distributed KAN training. The methodology is informed by established HPC benchmarking practice [14] while adapting the experimental protocol to the specifics of KAN's spline-parameterized layers, whose per-edge coefficients are updated jointly with the usual weight gradients during backpropagation.

The evaluation follows a systematic multi-dimensional design covering four aspects of parallel training: strong scaling, weak scaling, communication overhead, and model-size scaling. Strong-scaling experiments hold the total workload fixed while varying the number of GPUs, revealing how efficiently additional hardware is used. Weak-scaling experiments hold the per-GPU workload fixed while scaling both data and resources proportionally, revealing whether per-GPU performance remains stable as the system grows. Communication-overhead experiments isolate the cost of gradient synchronization across intra-node, inter-node, and hybrid topologies. Model-scaling experiments vary network width to relate parameter count to throughput, training time, and memory usage.

Experimental rigor is maintained by changing only one variable per experiment, allocating dedicated compute resources to avoid interference from co-located jobs, and including a system warm-up phase before timing measurements begin.

### 3.1. Performance Metrics

We report four primary scalability metrics. Speedup is defined as Sp = T1/Tp, where T1 is the single-GPU execution time and Tp is the execution time on p GPUs. Parallel efficiency is Ep = (Sp/p) × 100%, quantifying how effectively additional GPUs are used. Throughput is the number of samples processed per second, and per-GPU throughput normalizes total throughput by GPU count to expose per-device utilization trends. We additionally report communication overhead as the percentage of total step time spent on gradient synchronization, and the parameter-to-memory ratio – parameter count divided by peak GPU memory usage, a measure of parameter storage density rather than of memory utilization efficiency in the usual sense – to characterize how memory footprint scales with model size.

### 3.2. Experimental Platform

All experiments were conducted on the FinisTerrae III supercomputer at CESGA. Each compute node is equipped with 2× NVIDIA A100 GPUs, 2× Intel Xeon Ice Lake 8352Y processors (64 cores total), and 1 TB of DDR4 memory. Intra-node GPU communication uses PCIe 4.0; inter-node communication uses 200 Gb/s HDR InfiniBand in a fat-tree topology, with sub-microsecond intra-node and approximately 1–2 μs inter-node latency. The software stack consists of Rocky Linux 8.4, Python 3.9.7, and the SLURM scheduler, with NCCL as the primary collective communication backend. The system CUDA driver/toolkit installation is version 11.2, while PyTorch 2.2.2 was installed as a self-contained wheel bundling its own CUDA 12.1 runtime; the two coexist because PyTorch wheels ship the CUDA runtime libraries they were built against, independent of the system-level toolkit version, and the NVIDIA driver present on FinisTerrae III supports both.

Because each node provides two physical GPUs, some multi-node configurations in this study deliberately request only one GPU per node (via *--ntasks-per-node=1* and *--gres=gpu:a100:1*) rather than filling both GPUs on fewer nodes. This choice isolates pure inter-node communication from the intra-node PCIe effects that would otherwise be present whenever two ranks share a node. Configuration 4G4N (Section 4.1) uses this one-GPU-per-node allocation across four nodes, whereas 8G4N requests two GPUs per node (--*gres=gpu:a100:2*) across the same four nodes, so that the two configurations differ specifically in whether any GPU pair shares a node — the comparison exploited directly in the communication-overhead analysis of Section 4.3.

### 3.3. Model and Data

The model under test is a KAN with input dimension 2, one hidden layer of configurable width, and a scalar output, built on top of the reference PyKAN library [1] with a fixed spline order $k = 3$ and grid size 10. The spline order $k = 3$ (cubic splines) matches PyKAN's default; the grid size of 10 is a deliberate, moderately fine choice above PyKAN's default of 3, selected to keep spline resolution fixed and non-trivial across all model-scaling configurations in Section 4.4 rather than to explore grid size as an independent variable. This configuration deliberately uses the standard, non-accelerated recursive B-spline evaluation path – the same computational primitive targeted by operator-level optimizations such as MatrixKAN [5] – so that the reported communication and memory-scaling trends reflect DDP behavior under a representative, unoptimized KAN workload rather than a custom or simplified reimplementation.

Experiments use a synthetic two-dimensional regression task, $f(x, y) = x^2 + y^2$, with inputs drawn uniformly and controlled Gaussian noise added to the targets. Data is partitioned deterministically across workers with equal per-GPU sample counts and fixed random seeds for reproducibility. We adopt a synthetic task specifically so that dataset size, noise level, and per-sample compute cost can be held constant across all scaling configurations; Section 5.5 discusses the resulting limitation that these results have not yet been validated on real scientific or engineering datasets, where input dimensionality and function complexity differ substantially from this controlled setting.

### 3.4. Experimental Configurations

Strong-scaling experiments fix the total dataset at 100,000 samples while scaling from 1 GPU (1G1N) to 8 GPUs across 4 nodes (8G4N). Weak-scaling experiments fix the per-GPU workload at 25,000 samples, so that total dataset size grows proportionally with GPU count (from 25,000 samples on 1 GPU to 200,000 samples on 8 GPUs). Communication-overhead experiments isolate intra-node (PCIe), inter-node (InfiniBand), and hybrid communication patterns across matched GPU counts. Model-scaling experiments vary hidden width across four sizes – Small (32), Medium (64), Large (128), and Extra-Large (256) – corresponding to parameter counts from 2,726 to 21,542, under a fixed 2G4N distributed configuration.

Each configuration was trained once, for 40 epochs; the throughput and communication-overhead figures reported in Section 4 are means computed across all 40 epochs of that run, and Sections 4.1, 4.3, and 4.4 additionally report the standard deviation and coefficient of variation across those same 40 epochs, which capture within-run measurement variability – including early-epoch warm-up effects – rather than run-to-run variability from independently repeated training. This is distinct from the coefficient of variation reported in Section 4.2, which is computed across the different GPU configurations themselves, not across epochs of a single configuration. The two also use different standard-deviation conventions, chosen to match what each is estimating: Section 4.2 treats its six configurations as the complete set of interest and uses the population standard deviation (dividing by $n$), whereas the per-epoch $\sigma$ and CV in Tables 1, 3, and 4 treat the 40 epochs as a sample drawn from the training run and use the sample standard deviation (dividing by $n - 1$), the standard convention for estimating variability from a finite sample.

## 4. Experimental Results

This section presents results across the four evaluation dimensions introduced in Section 3: strong scaling (Section 4.1), weak scaling (Section 4.2), communication overhead (Section 4.3), and model-size scaling (Section 4.4).

## 4.1. Strong Scaling Performance Analysis

Strong scaling examines how effectively a fixed workload of 100,000 samples can be distributed across an increasing number of GPUs, from single-GPU baseline to an 8-GPU, 4-node configuration.

### *4.1.1. Speedup and Parallel Efficiency*

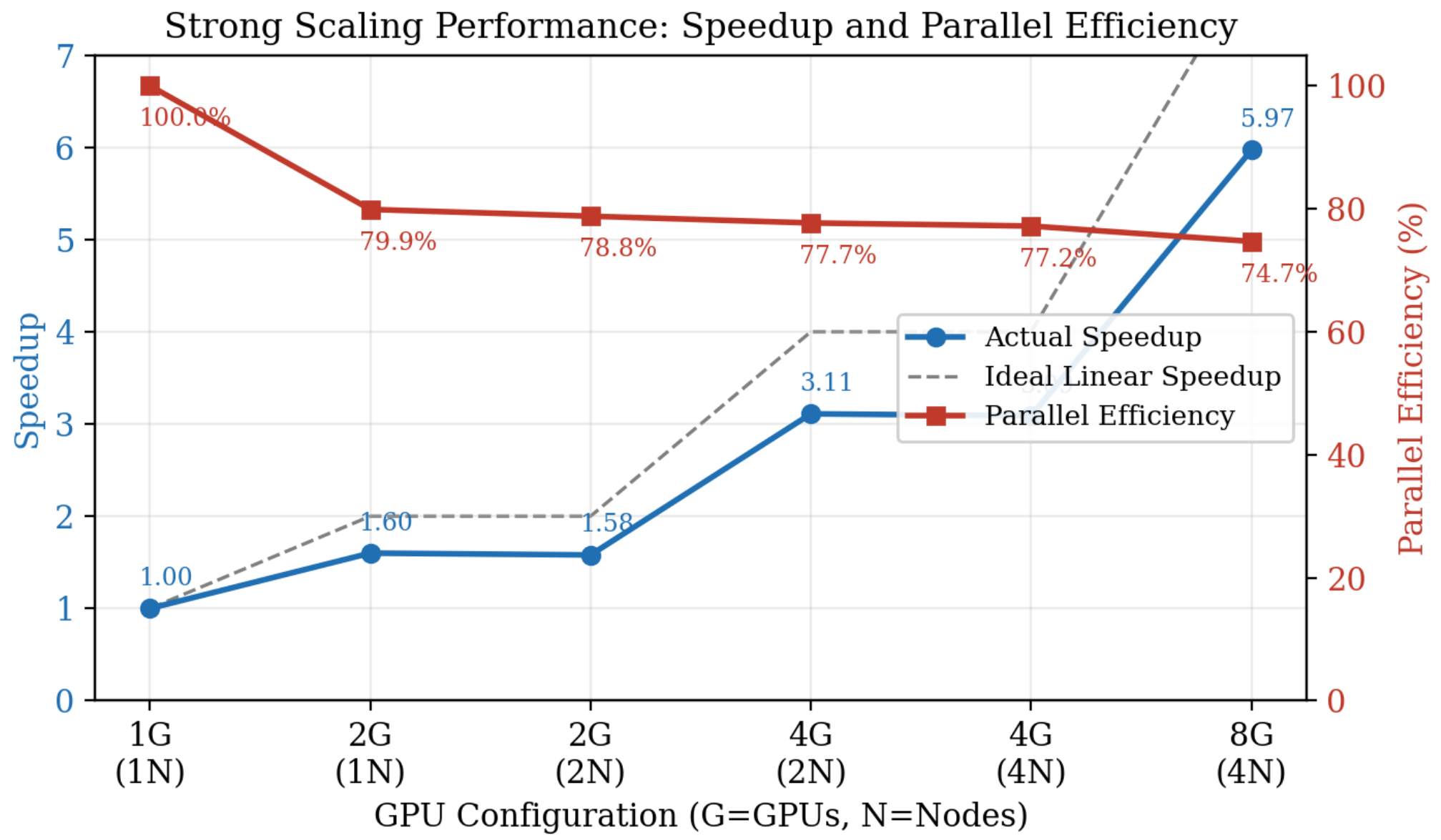

**Figure 2: Strong scaling speedup and parallel efficiency.**

**Table 1: Strong scaling performance results.**

| Config | GPUs | Time (s) | Throughput (samples/s) | σ | CV (%) | Speedup | Parallel Eff. (%) |
|---|---|---|---|---|---|---|---|
| 1G1N | 1 | 900.59 | 4,442 | ±38 | 0.9 | 1.00 | 100.0 |
| 2G1N | 2 | 563.58 | 7,099 | ±67 | 0.9 | 1.60 | 79.9 |
| 2G2N | 2 | 571.73 | 6,998 | ±88 | 1.3 | 1.58 | 78.8 |
| 4G2N | 4 | 289.73 | 13,819 | ±307 | 2.2 | 3.11 | 77.7 |
| 4G4N | 4 | 291.50 | 13,734 | ±294 | 2.1 | 3.09 | 77.2 |
| 8G4N | 8 | 150.80 | 26,618 | ±1212 | 4.6 | 5.97 | 74.7 |

*σ is the standard deviation of the reported quantity across all 40 training epochs of the (single) run for that configuration, matching the window over which the reported mean is computed; it reflects within-run measurement variability rather than variability across independently repeated training runs. CV% is σ as a percentage of the mean, to make magnitudes comparable across configurations with different absolute throughput. Throughput σ values (Tables 1 and 4) are substantially inflated by a single early-epoch warm-up effect – the first epoch after process-group initialization is consistently slower – and fall markedly (in most configurations to well under half) if that epoch is excluded; communication-overhead σ values (Table 3) are not: excluding the first epoch changes them by at most 9%, indicating that the overhead itself fluctuates throughout training rather than only at startup.*

Figure 2 presents speedup and parallel efficiency across GPU configurations. KAN training reaches a speedup of 5.97× at 8 GPUs relative to the single-GPU baseline, corresponding to a parallel efficiency of 74.7%. Efficiency declines gradually and monotonically with GPU count – 100.0%, 79.9%, 78.8%, 77.7%, 77.2%, and 74.7% for the 1, 2 (intra-node), 2 (inter-node), 4 (2-node), 4 (4-node), and 8-GPU configurations, respectively (Table 1).

This efficiency range (74.7%–100%) is broadly consistent with DDP overhead patterns reported for conventional MLP and CNN workloads, where gradient synchronization and All-Reduce latency are known to erode parallel efficiency below the ideal linear bound as GPU count grows [7, 10]. We do not observe evidence that KAN's spline-based, per-edge activation parameters introduce an additional scaling penalty beyond what is expected from standard DDP overhead. Each spline basis-function evaluation and its gradient are local to a single edge and independent across edges and samples within a layer, so the added computation is plausibly close to embarrassingly parallel. It would therefore not be expected to introduce extra synchronization points beyond the single gradient All-Reduce already required by DDP. The practical implication is that KAN's higher per-sample compute cost, relative to an MLP with a similar node count, changes the compute-to-communication ratio (Section 5.4) but does not change the qualitative shape of the scaling curve.

A notable pattern is that node distribution has a measurable but secondary effect at fixed GPU count: 2G2N (two GPUs on separate nodes) underperforms 2G1N (two GPUs on one node) by roughly one percentage point of efficiency, while 4G4N and 4G2N are nearly identical. Section 4.3 examines this pattern directly through per-topology communication-overhead measurements and traces it to All-Reduce algorithm selection rather than to any KAN-specific effect.

The observed 8-GPU speedup implies, via Amdahl's law, an effective non-parallelizable fraction of approximately 4.9% – which should be read as an Amdahl-equivalent summary of all overheads that do not shrink with GPU count (communication, synchronization, and any residual load imbalance), rather than as a literal serial code fraction – a plausible order of magnitude for fixed per-step DDP overhead such as gradient All-Reduce initiation and synchronization barriers.

### *4.1.2. Throughput Analysis*

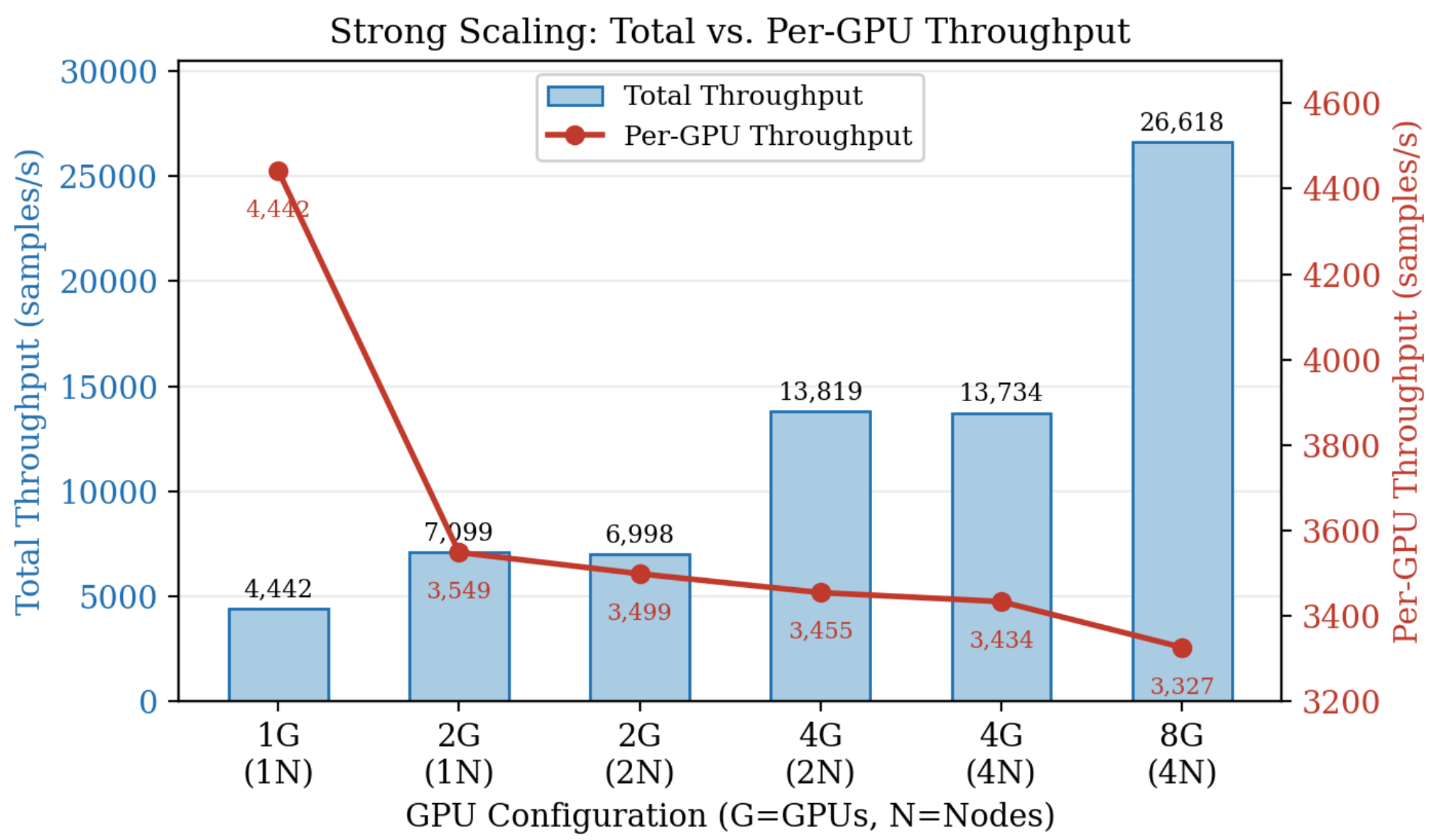

**Figure 3: Strong scaling: total vs. per-GPU throughput.**

Figure 3 shows total and per-GPU throughput across the same configurations. Total throughput scales from 4,442 samples/s on 1 GPU to 26,618 samples/s on 8 GPUs – a 5.99× increase that closely tracks the measured speedup, as expected since both are computed from the same wall-clock measurements. Per-GPU throughput declines from 4,442 to 3,327 samples/s/GPU (a 25.1% reduction) over the same range, mirroring the efficiency decline in Table 1 and quantifying the same phenomenon from a resource-utilization perspective: each additional GPU contributes less incremental throughput than the previous one, at a rate set by growing communication overhead rather than by any change in per-sample computational cost.

These results indicate that, for the workload and model sizes evaluated here, 8 GPUs remain within the practically useful scaling range for KAN training under DDP, with further scaling expected to yield diminishing but still positive returns for sufficiently large workloads, consistent with the general behavior predicted by Amdahl's law for a workload with a small non-parallelizable fraction – as noted above, on the order of 4.9% here.

## 4.2. Weak Scaling Stability Analysis

Weak scaling holds the per-GPU workload fixed at 25,000 samples while both the total dataset size and GPU count grow proportionally, isolating whether per-GPU performance is preserved as the system scales.

### *4.2.1. Per-GPU Throughput Stability*

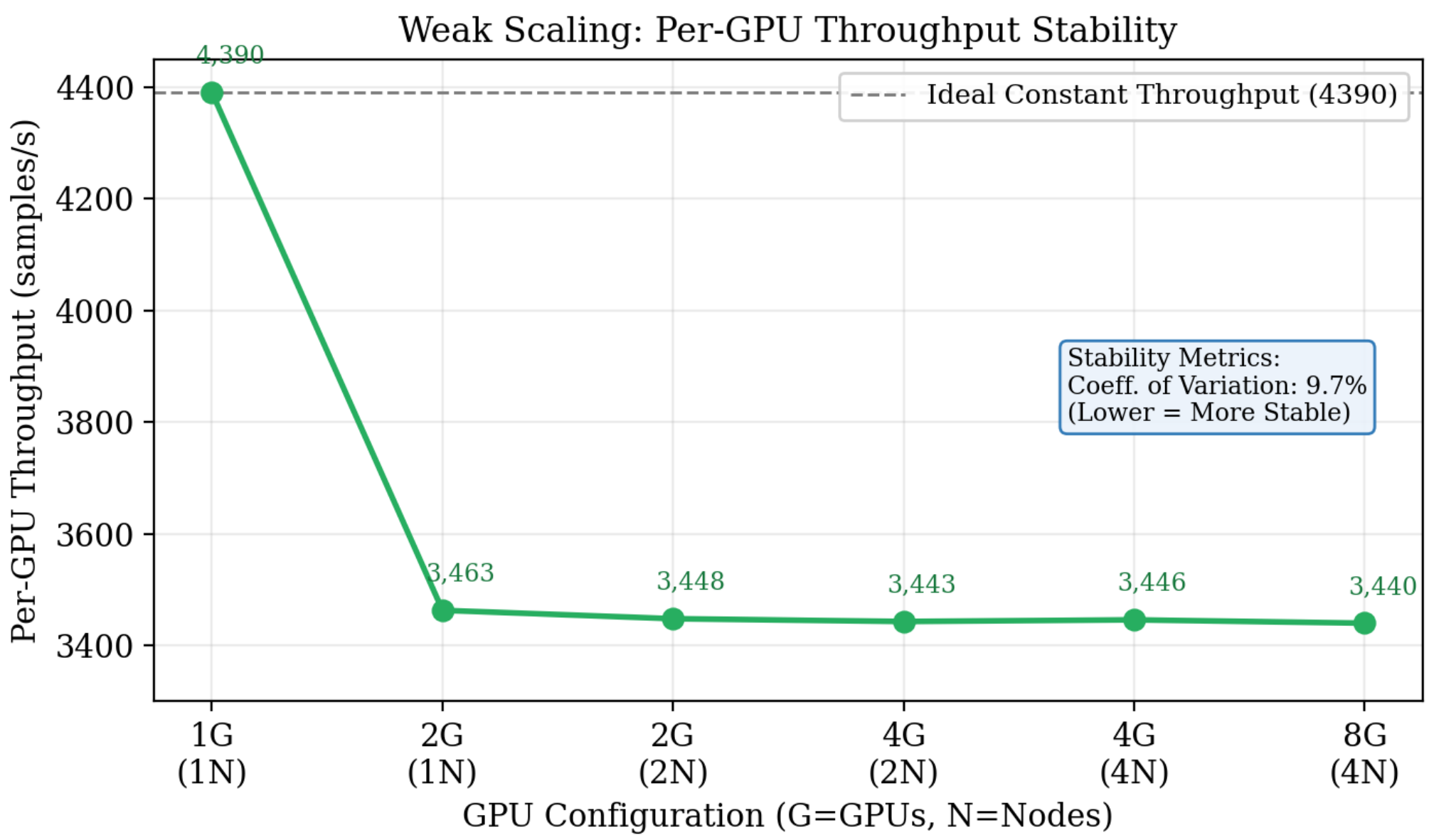

**Figure 4: Weak scaling: per-GPU throughput stability.**

**Table 2: Weak scaling performance results.**

| Config | GPUs | Throughput (samples/s/GPU) | Time (s) |
|---|---|---|---|
| 1G1N | 1 | 4,390 | 227.94 |
| 2G1N | 2 | 3,463 | 288.98 |
| 2G2N | 2 | 3,448 | 290.39 |
| 4G2N | 4 | 3,443 | 290.63 |
| 4G4N | 4 | 3,446 | 290.35 |
| 8G4N | 8 | 3,440 | 290.98 |

Figure 4 and Table 2 show per-GPU throughput across configurations. The single-GPU baseline (1G1N) reaches 4,390 samples/s/GPU, while every distributed configuration – from 2 to 8 GPUs – clusters tightly between 3,440 and 3,463 samples/s/GPU. The coefficient of variation is 9.7% across all six configurations, falling to under 0.3% among the five distributed configurations alone.

The single-GPU configuration is not directly comparable to the distributed configurations because it operates without any DDP overhead: no gradient All-Reduce, no NCCL process-group initialization, and no cross-process synchronization barrier. The approximately 21% single-to-multi-GPU throughput drop therefore reflects a largely fixed per-step DDP overhead rather than a scaling deficiency – and, importantly, this overhead does not grow further as GPU count increases from 2 to 8. The near-constant wall-clock time of approximately 290 seconds across all distributed configurations (Table 2) confirms that the added communication introduced by larger GPU counts is fully amortized by the proportionally larger per-run dataset, which is the defining signature of good weak scaling.

This stability is favorable for practical deployment: it indicates that, once the fixed cost of entering distributed training is paid, KAN's per-GPU efficiency does not further degrade as the training job is scaled out to more workers, provided the per-GPU workload is kept approximately constant. Because this fixed-overhead behavior is characteristic of DDP training generally, rather than of KAN's spline-based computation specifically, we treat it as evidence that KAN does not introduce distinguishing weak-scaling behavior relative to conventional architectures under this workload regime, and we do not draw a stronger architecture-specific conclusion from it.

### 4.3. Communication Overhead Analysis

Communication overhead is the fraction of total step time spent synchronizing gradients across GPUs, and it is the primary driver of the parallel-efficiency decline observed in Section 4.1. Table 3 and Figure 5 break this overhead down by topology.

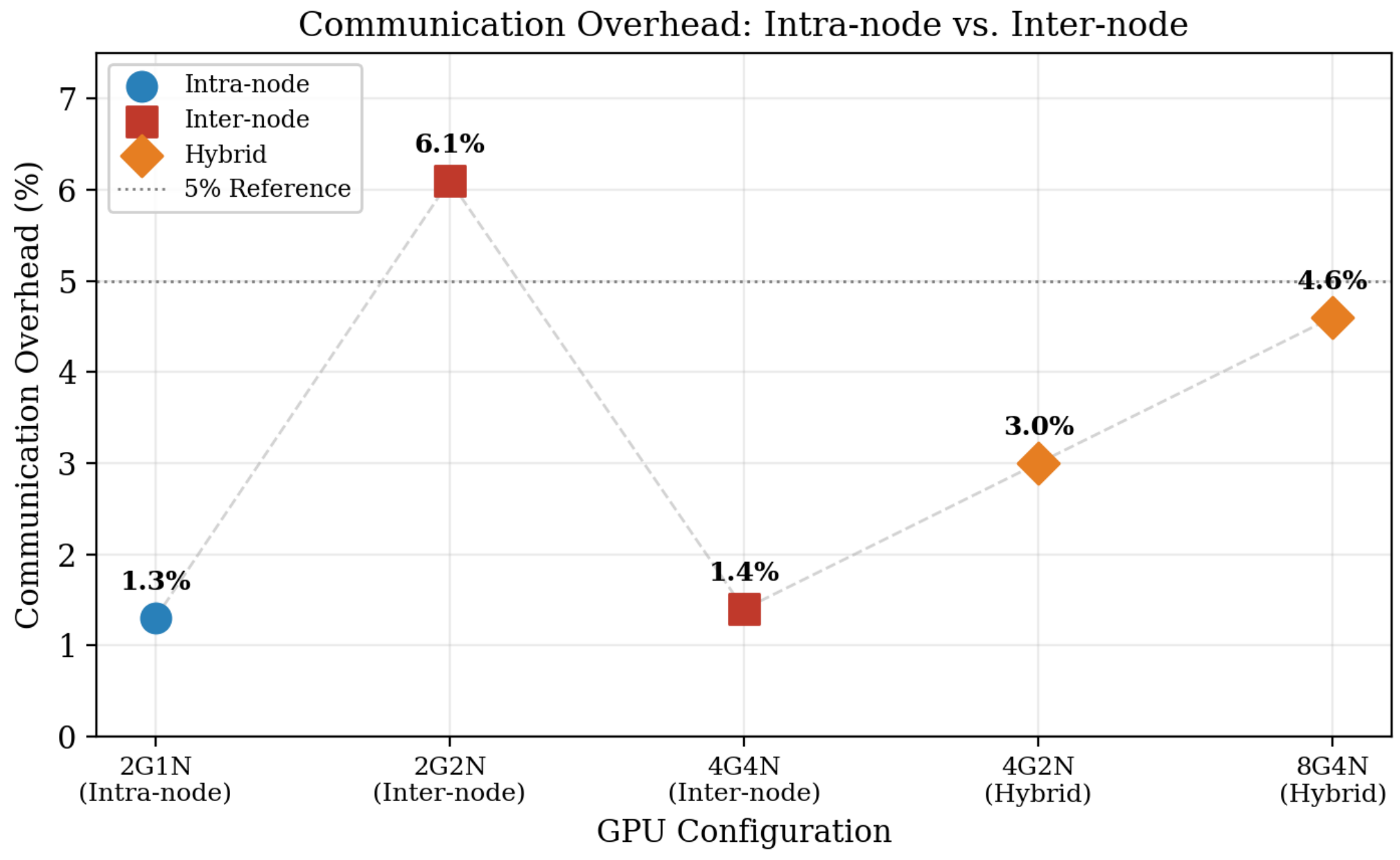

**Figure 5: Communication overhead across intra-node, inter-node, and hybrid GPU topologies.**

**Table 3: Communication overhead by configuration.**

| Config | GPUs | Type | Overhead (%) | σ | CV (%) | Efficiency (%) |
|---|---|---|---|---|---|---|
| 2G1N | 2 | Intra-node | 1.3 | ±0.05 | 4.2 | 98.7 |
| 2G2N | 2 | Inter-node | 6.1 | ±0.81 | 13.3 | 93.9 |
| 4G4N | 4 | Inter-node | 1.4 | ±0.40 | 28.9 | 98.6 |
| 4G2N | 4 | Hybrid | 3.0 | ±1.25 | 41.8 | 97.0 |
| 8G4N | 8 | Hybrid | 4.6 | ±0.69 | 15.1 | 95.4 |

The overhead pattern is non-monotonic in GPU count – it rises sharply from 2G1N (1.3%) to 2G2N (6.1%), falls to 4G4N (1.4%), then rises again through 4G2N (3.0%) to 8G4N (4.6%). We interpret this pattern mechanistically, as a consequence of All-Reduce algorithm behavior and network topology rather than of any property specific to KAN's gradient structure.

Two-GPU inter-node penalty. For two-rank cases, NCCL's All-Reduce may reduce to a simple pairwise exchange: each GPU sends its full gradient to its peer and waits for the reciprocal transfer, with no opportunity for pipelining across a ring or tree. However, we did not capture NCCL_DEBUG traces to confirm the specific algorithm selected, so this mechanism is inferred from the overhead pattern rather than directly observed. Under this hypothesis, when both ranks are inter-node (2G2N), the pairwise exchange must additionally cross the InfiniBand fabric. The fixed per-message latency of that single cross-node link would then dominate total communication time, since there is no second link to overlap it with. This provides a plausible explanation for why 2G2N (6.1%) is markedly worse than 2G1N (1.3%), where the same exchange pattern would instead stay on the lower-latency, higher-bandwidth PCIe intra-node link.

Four-GPU ring advantage. At four ranks, NCCL typically has the option to construct a ring or tree All-Reduce, in which gradient chunks are pipelined across multiple segments simultaneously rather than exchanged pairwise. A structural change of this kind, rather than a KAN-specific effect, plausibly explains why 4G4N reaches 1.4% overhead despite using exclusively inter-node links (recall from Section 3.2 that 4G4N deliberately allocates one GPU per node across four nodes): with four uniformly-placed inter-node connections, a ring-style algorithm would overlap communication across segments and better utilize the available InfiniBand bandwidth than the two-rank case allows.

Hybrid-topology penalty at 4G2N and 8G4N. When multiple GPUs share a node (4G2N: two nodes with two GPUs each; 8G4N: four nodes with two GPUs each), NCCL likely needs to coordinate a two-tier hierarchy, rather than a single uniform ring: an intra-node reduction over PCIe, followed by an inter-node reduction over InfiniBand. 4G2N (3.0%) has higher overhead than 4G4N (1.4%) despite using the same total GPU count and, if anything, more favorable intra-node bandwidth. This indicates that the coordination cost of synchronizing between two communication tiers with different bandwidth/latency characteristics outweighs the benefit of the faster intra-node link in this regime. The same hierarchical-coordination cost, compounded across four nodes with two tiers each, is consistent with the further increase to 4.6% overhead observed at 8G4N.

Overall, efficiency stays above 95% in four of the five configurations, with 2G2N as the clear outlier. Two practical implications follow directly: first, when only two GPUs are available, colocating them on a single node is preferable to splitting them across nodes; and second, when scaling beyond two GPUs, topologies that avoid mixed intra-/inter-node tiers (one GPU per node, as in 4G4N) can outperform topologies that mix them (two GPUs per node, as in 4G2N or 8G4N), even though the latter nominally offer faster intra-node links. We return to these observations as concrete configuration guidance in Section 5.3.

We interpret these mechanisms as unlikely to be specific to KAN: they are properties of NCCL's All-Reduce algorithm selection and of the InfiniBand/PCIe latency hierarchy, and the same qualitative pattern would plausibly be expected for any DDP-trained model with a comparable gradient tensor size. This interpretation follows from established distributed-systems principles rather than from a matched-architecture ablation; Section 5.5 discusses this limitation directly. What is KAN-specific is only the magnitude of the gradient tensor being synchronized – since KAN's per-edge spline coefficients add to the parameter count relative to an MLP of the same node topology – and Section 4.4 shows that this magnitude effect becomes more significant as model width increases.

## 4.4. Model Scaling Characteristics Analysis

This section examines how KAN performance changes with model width, under a fixed 2G4N distributed configuration, across four sizes: Small (width 32, 2,726 parameters), Medium (64, 5,414), Large (128, 10,790), and Extra-Large (256, 21,542).

### *4.4.1. Performance-Parameter Trade-off*

**Table 4: Model scaling performance results.**

| Size | Width | Parameters | Throughput (samples/s) | σ | CV (%) | Memory (GB) |
|---|---|---|---|---|---|---|
| Small | 32 | 2,726 | 24,624 | ±1029 | 4.2 | 0.04 |
| Medium | 64 | 5,414 | 13,344 | ±417 | 3.1 | 0.06 |
| Large | 128 | 10,790 | 6,867 | ±85 | 1.2 | 0.10 |
| X-Large | 256 | 21,542 | 3,469 | ±35 | 1.0 | 0.18 |

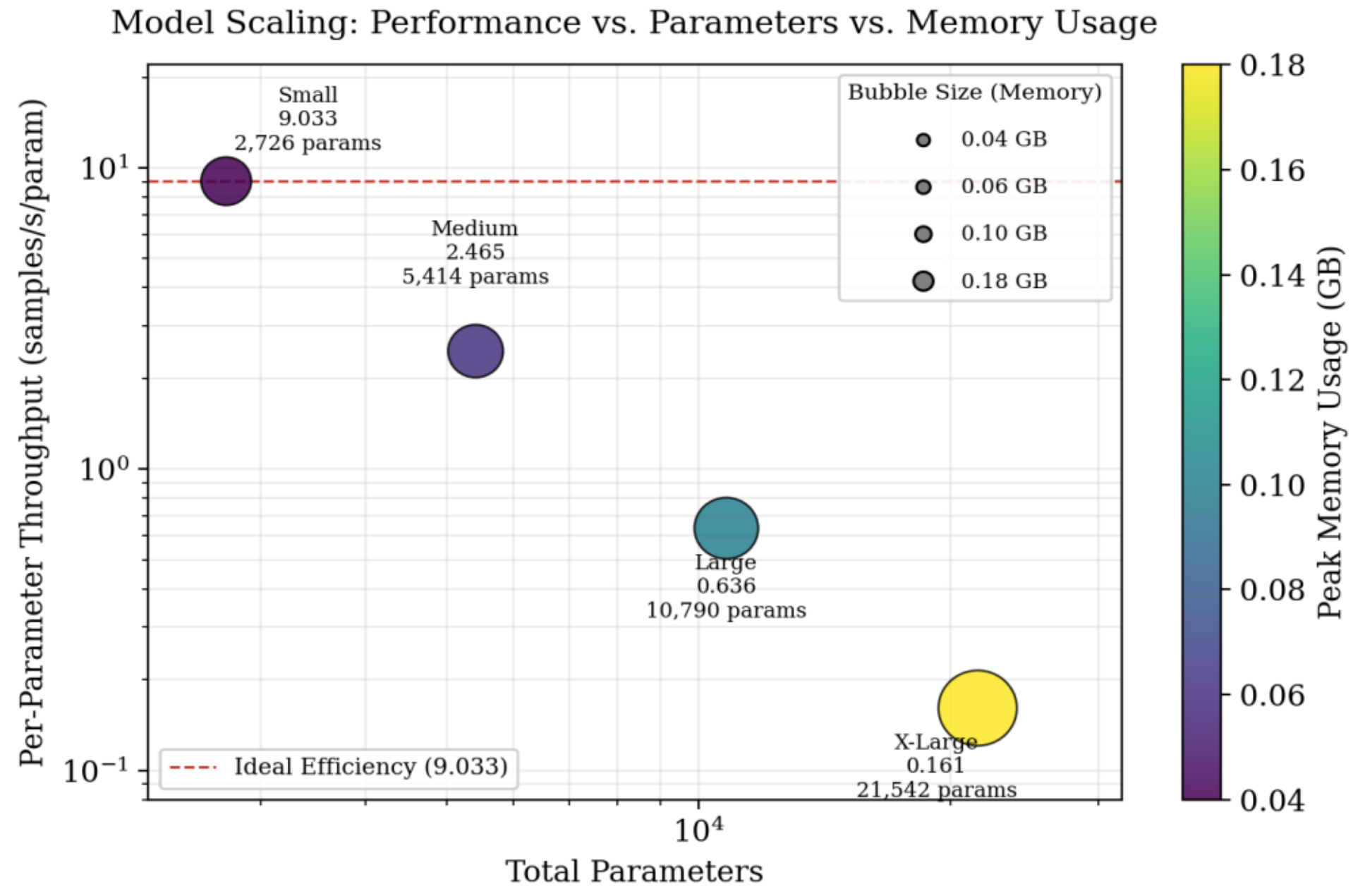

**Figure 6: Model scaling: throughput vs. parameters vs. memory usage.**

Table 4 and Figure 6 summarize throughput, training time, and per-parameter efficiency across model sizes. Training time grows from 81.5 s (Small) to 577.0 s (Extra-Large) as width increases 8×, a 7.08× increase in wall-clock time. A strictly constant per-edge cost combined with a linear increase in edge count would predict exactly an 8× growth in training time; the observed 7.08× is modestly below this, a gap of about 11%. We do not have a confirmed explanation for this specific gap – plausible contributors include partial amortization of fixed per-step overhead (data loading, optimizer step, communication) over a larger per-step compute cost as width grows, or minor deviations from perfectly uniform per-edge cost across grid positions – but in either case the result is close enough to linear, and sufficiently far from constant, to support the qualitative conclusion that training time scales approximately linearly with width rather than remaining flat, as a purely embarrassingly-parallel computation with fixed per-sample cost would imply. Correspondingly, total throughput falls from 24,624 to 3,469 samples/s (a 7.1× decrease), and per-parameter throughput falls 56×, from 9.033 to 0.161 samples/s/parameter.

The near-linear scaling of training time with width is attributable to the structure of B-spline evaluation itself: each edge activation requires evaluating k+1 overlapping basis functions and accumulating their contribution through the recursive Cox-de Boor formulation used by the reference PyKAN implementation, and both the forward evaluation and the backward gradient computation scale with grid size and spline order per edge, in addition to scaling with the number of edges. Doubling network width increases the edge count linearly but does not change the per-edge basis-function cost, so the observed trend is consistent with per-edge cost remaining roughly constant while the number of edges scales with width – rather than reflecting genuinely superlinear (e.g., quadratic) per-edge computational complexity; distinguishing between these explanations precisely, and specifically accounting for the 11% gap noted above, would require kernel-level profiling, which we did not perform and which is the province of operator-level work such as MatrixKAN [5], rather than of the present system-level study.

Model quality improves with scale: final training loss decreases from 0.100 (Small) to 0.080 (Extra-Large), though the improvement is modest relative to the 56× per-parameter efficiency cost, indicating diminishing accuracy returns from width alone on this two-dimensional regression task.

#### *4.4.2. Parameter-to-Memory Ratio Trends*

The same measurements (Figure 6, bubble size and color encode peak memory) reveal a counter-intuitive result: the parameter-to-memory ratio, defined as parameter count divided by peak GPU memory usage, improves with model size, from 74,268 parameters/GB (Small) to 118,066 parameters/GB (Extra-Large) – despite peak memory usage itself growing from 0.04 GB to 0.18 GB.

We hypothesize that this trend reflects fixed per-process memory overhead being amortized over a larger parameter set as width increases. Distributed training carries memory costs that plausibly do not scale with model size: NCCL communication buffers, the DDP process's CUDA context, PyTorch's caching allocator overhead, and pinned-memory staging buffers. At the Small configuration (2,726 parameters, 0.04 GB total), these largely fixed costs would make up a proportionally larger share of the 0.04 GB footprint than they would of the Extra-Large configuration's 0.18 GB footprint. Under this hypothesis, the amortization effect would again be a general property of the distributed training infrastructure rather than of KAN's spline parameterization specifically; the KAN-specific contribution would be only that spline coefficients scale sub-linearly with reported parameter count in terms of raw storage (each coefficient is a single float, no different from an MLP weight), so we would expect a similar amortization pattern, with a comparable shape, for any small-to-moderate DDP-trained model measured across a comparable size range. We did not, however, isolate the fixed-overhead and parameter-storage components directly in these experiments, and doing so – for instance by measuring memory usage with communication disabled, or by repeating this sweep for a matched-size MLP – remains necessary to confirm this explanation quantitatively rather than treat it as established.

## 5. Comprehensive Performance Evaluation and Discussion

### 5.1. Multi-Dimensional Performance Synthesis

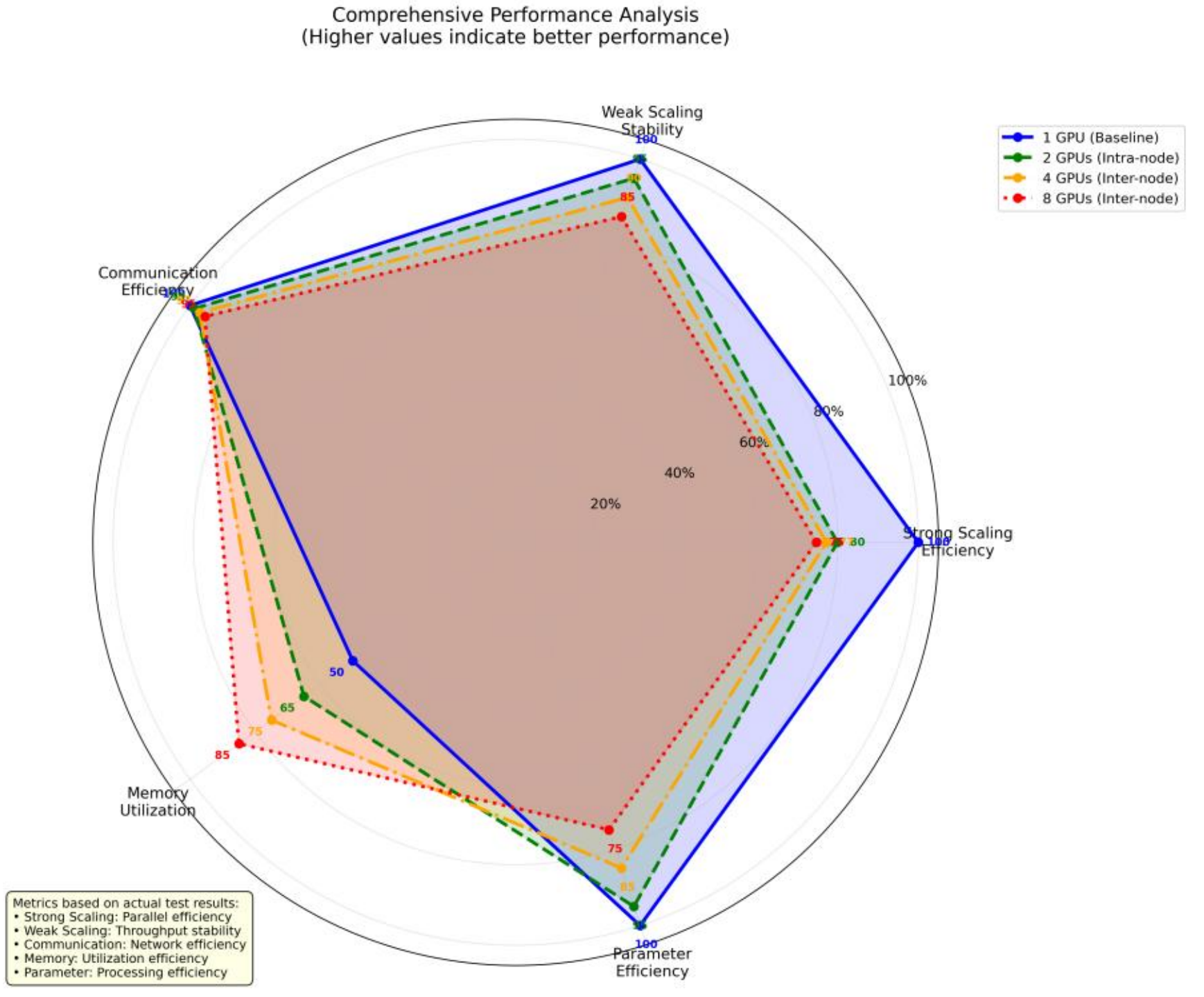

**Figure 7: Comprehensive multi-dimensional performance comparison across GPU configurations.**

Figure 7 synthesizes the four evaluation dimensions – compute efficiency, communication efficiency, memory utilization, and scalability – across representative GPU configurations, each normalized to a 0–100 scale. Because these dimensions are not measured in commensurate units, the normalized scores should be read as a qualitative visual summary of the trade-offs identified in Sections 4.1–4.4, rather than as a single quantitative composite metric.

No single configuration dominates on all four axes simultaneously, which is expected given the mechanisms identified in Sections 4.1–4.4: the single-GPU configuration maximizes compute efficiency by construction (no communication overhead exists to measure), but offers no scalability benefit; the 4-GPU configurations offer the best balance of scalability and communication efficiency, consistent with the ring-based All-Reduce advantage identified in Section 4.3; and the 8-GPU configuration trades some communication efficiency and parameter-to-memory ratio for the highest absolute throughput. This trade-off structure indicates that the appropriate GPU configuration for a given deployment depends on whether the workload is throughput-bound (favoring 8 GPUs) or efficiency-bound (favoring 4 GPUs), a distinction we return to in Section 5.3.

## 5.2. Optimization Strategies

The mechanisms identified in Sections 4.3 and 4.4 point to concrete optimization directions, some of which are general to DDP training and some of which are specific to KAN's parameter structure.

Communication. The two-tier hybrid-topology penalty identified at 4G2N and 8G4N (Section 4.3) suggests that a hierarchical All-Reduce – aggregating within a node over PCIe first, then across nodes over InfiniBand – could reduce coordination overhead relative to treating all ranks uniformly, particularly as node count grows. Overlapping gradient communication with backward-pass computation (gradient bucketing, as already implemented in PyTorch DDP) is also likely to help, since KAN's per-edge gradient computation is itself parallelizable and could proceed concurrently with the All-Reduce of earlier layers' gradients.

Computation. The growth in training time with model width (Section 4.4) motivates operator-level optimization of the spline evaluation itself. This is precisely the problem addressed by MatrixKAN's matrix-formulated B-spline evaluation [5], which we did not implement or evaluate in this study but which is directly complementary to the data-parallel setting studied here: replacing the reference recursive spline evaluation with a matrix-batched one inside the same DDP training loop should reduce per-GPU compute time without changing the communication-overhead mechanisms identified in Section 4.3, and, as noted in Section 2.2, would make communication-aware topology selection more important in relative terms, since a smaller compute time increases the fraction of step time spent on communication for a fixed communication cost.

## 5.3. Deployment Guidance

Based on the mechanisms identified above, we offer the following configuration guidance, restricted to the workload and hardware regime studied here (two-dimensional regression, models up to 21.5K parameters, A100 GPUs with HDR InfiniBand).

GPU topology. When only two GPUs are available, colocating them on a single node avoids the pairwise-exchange, cross-node latency penalty identified in Section 4.3 (1.3% vs. 6.1% overhead). At four GPUs, distributing one GPU per node (4G4N) is preferable to two GPUs per node (4G2N), since it avoids the two-tier communication-hierarchy coordination cost while still benefiting from ring-based All-Reduce. Beyond four GPUs, further scaling to eight GPUs remains beneficial for throughput-bound workloads but introduces additional hybrid-topology overhead (Section 4.3) that should be weighed against the specific workload's tolerance for reduced efficiency.

Model size. Because per-parameter throughput falls sharply with width while accuracy gains are comparatively modest on the regression task studied here (Section 4.4.1), we recommend selecting model width incrementally, validating that each width increase yields a favorable accuracy-to-compute-cost trade-off for the specific application, rather than defaulting to the largest model the memory budget allows. The parameter-to-memory ratio does improve with width (Section 4.4.2), so memory pressure alone is not a strong argument against larger models; the binding constraint is more often training-time budget.

## 5.4. Complementarity with Operator-Level Optimization

We return here to the relationship between this study and MatrixKAN [5] introduced in Section 2.2, since it is central to correctly scoping our contribution. This paper does not propose a faster KAN operator, and it would be inaccurate to describe our findings as an alternative to MatrixKAN's approach. Rather, the two contributions sit at different layers of the training stack: MatrixKAN reduces Tcompute for a single forward/backward pass; this study characterizes how Tcomm scales with GPU count and topology once that pass is replicated across devices. A production KAN training system operating at scale would plausibly use both: an operator-level acceleration such as MatrixKAN to reduce per-GPU step time, combined with the topology-aware DDP configuration guidance derived here to keep the resulting (proportionally larger) communication overhead in check. Quantifying this combined effect directly – for instance, by re-running the strong-scaling experiments of Section 4.1 with a MatrixKAN-accelerated model – is a direct and, we believe, high-value extension of this work that we did not perform here.

## 5.5. Limitations and Future Directions

Several limitations bound the generality of these findings and should inform how they are used.

Synthetic workload. All experiments use a two-dimensional synthetic regression task ($f(x, y) = x^2 + y^2$). This choice was deliberate, allowing precise control over dataset size and per-sample cost across scaling configurations, but it means the reported throughput and communication-overhead numbers have not been validated on higher-dimensional or real scientific/engineering datasets, where input dimensionality, spline grid resolution, and batch composition may differ substantially and could shift the compute-to-communication ratio underlying the scaling curves in Sections 4.1–4.3.

Model scale. The largest model evaluated has 21,542 parameters, several orders of magnitude smaller than KAN deployments reported in scientific-computing applications [4]. The parameter-to-memory ratio trend identified in Section 4.4.2 is extrapolated from a narrow parameter range, and whether it continues, plateaus, or reverses at much larger scale is untested. This range was not extended further due to the fixed timeline of the MSc thesis project underlying this study; access to the FinisTerrae III allocation ended at the project's conclusion, not because of any memory or compute limit encountered at the largest width evaluated.

Single cluster, moderate GPU count. All experiments were conducted on one HPC system (FinisTerrae III) at up to 8 GPUs across 4 nodes. The communication-overhead mechanisms identified in Section 4.3 are grounded in general NCCL/InfiniBand behavior and are likely to generalize qualitatively to other fat-tree InfiniBand clusters, but the specific overhead percentages reported here are platform-specific and should not be read as universal constants.

Reference (unoptimized) spline evaluation. As discussed in Section 5.4, we evaluate the standard recursive B-spline computation path rather than an operator-level-accelerated one. The absolute training-time numbers in Section 4.4 are therefore upper bounds on what an optimized implementation would require; the qualitative scaling trends (near-linear time growth, sub-linear memory growth) are expected to persist under an accelerated operator, but this has not been empirically confirmed.

Non-KAN-specific mechanisms not directly isolated. Several of the mechanistic explanations offered in Sections 4.1–4.4 – fixed DDP overhead amortization, All-Reduce algorithm behavior, two-tier communication coordination cost – are argued from established distributed-systems principles and from the pattern of the measurements themselves, but were not confirmed through direct ablation. We regard this as the most significant limitation of the present study, since it is what would let a matched-architecture comparison distinguish general DDP behavior from anything specific to KAN's per-edge parameterization. We were unable to carry out this comparison within the present work because access to the FinisTerrae III allocation ended with the conclusion of the MSc thesis project underlying this paper, and we did not have the opportunity to run additional experiments before submission. We specify the design here so that it can be executed directly, by us or by others, without requiring further methodological work. An MLP would be constructed with the same input dimension (2) and the same number of hidden layers as the KAN models of Section 4.4, with a hidden width chosen so that its total parameter count matches each of the four KAN configurations to within a few percent (approximately 2.7K, 5.4K, 10.8K, and 21.5K parameters). This MLP would be trained under the identical strong-scaling protocol of Section 4.1, on the same two topology pairs already identified as diagnostic in Section 4.3 – 2G1N versus 2G2N, to isolate the intra-node/inter-node pairwise-exchange penalty, and 4G4N versus 4G2N, to isolate the two-tier hybrid-topology coordination cost – using the same synthetic regression task, dataset size, and number of repeated runs as the corresponding KAN experiments. If the MLP control reproduces the same qualitative pattern (2G2N overhead materially exceeding 2G1N; 4G2N overhead materially exceeding 4G4N) at closely comparable magnitudes, this would confirm that the mechanisms identified in Section 4.3 are properties of DDP and NCCL rather than of KAN's edge-wise gradient structure; a materially different pattern or magnitude would instead indicate a KAN-specific contribution to communication overhead that the present study has not isolated. Such a matched-architecture comparison would substantially strengthen the claim that these mechanisms are general rather than KAN-specific, and we identify it as the most direct next step for this line of work.

Building directly on the limitations above, we identify four priorities for follow-up work: (1) a matched-architecture control experiment comparing KAN and MLP scaling under identical DDP configurations, to directly test which of the mechanisms identified in Sections 4.1–4.4 are architecture-general versus KAN-specific; (2) validation on real scientific or engineering regression/PDE-surrogate datasets, where KAN has shown particular promise [4]; (3) combining this study's topology guidance with operator-level acceleration such as MatrixKAN [5], to quantify the compound effect discussed in Section 5.4; and (4) extending the evaluated scale beyond 8 GPUs and beyond 21.5K parameters, where memory and communication trends may behave differently.

# 6. Conclusions

This paper presented an empirical scalability study of data-parallel Kolmogorov-Arnold Network training under PyTorch DDP on multi-node, multi-GPU HPC infrastructure. Across four evaluation dimensions – strong scaling, weak scaling, communication overhead, and model-size scaling – KAN training reaches 74.7% parallel efficiency and a 5.97× speedup at 8 GPUs, exhibits stable weak-scaling behavior (9.7% coefficient of variation overall, 0.3% among distributed configurations) once the fixed cost of entering distributed training is amortized, and shows a non-monotonic but explicable communication-overhead pattern (1.3%–6.1%) driven by All-Reduce algorithm selection and node topology rather than by KAN's architecture per se. Larger models achieve a more favorable parameter-to-memory ratio (74K→118K parameters/GB) even as training time grows with width faster than a fixed-cost model would predict, reflecting the cost of per-edge B-spline evaluation.

Throughout, we have distinguished between findings that appear to be general properties of DDP-based distributed training – and would likely be observed for a comparably-sized MLP – and findings that are more plausibly specific to KAN's per-edge spline parameterization, principally the elevated communication volume and the growth in compute cost with width. We have positioned this contribution explicitly relative to concurrent operator-level KAN parallelization work such as MatrixKAN [5], arguing that the two directions address different layers of the training stack and are complementary rather than competing.

The results support the practical conclusion that KAN can be trained under standard data-parallel DDP without encountering distributed-training pathologies beyond those expected for conventional architectures, subject to the limitations discussed in Section 5.5 – most importantly, that these findings derive from a synthetic two-dimensional regression task and model sizes up to 21.5K parameters on a single HPC platform, and that validation on real workloads, larger models, and matched-architecture controls remains necessary before generalizing further. Our experimental scripts, logs, and analysis code are publicly available at https://github.com/anhushow/KAN_Scalability_DDP to support such follow-up work.

## CRediT authorship contribution statement

**Guangneng Chen**: Conceptualization, Methodology, Software, Validation, Formal analysis, Investigation, Data curation, Visualization, Writing – original draft, Writing – review & editing. **David García-Selfa**: Conceptualization, Supervision, Writing – review & editing. **Pablo Quesada Barriuso**: Supervision, Writing – review & editing.

## Declaration of competing interest

The authors declare that there is no known competing financial interest or personal relationship that could have appeared to influence the work reported in this paper.

## Funding

This research did not receive any specific grant from funding agencies in the public, commercial, or not-for-profit sectors.

## Data availability

The code and scripts used to run the experiments and generate the figures in this paper are publicly available at https://github.com/anhushow/KAN_Scalability_DDP.

## Acknowledgements

This paper is based on research carried out for the first author's MSc thesis at the University of Santiago de Compostela, supervised by Pablo Quesada Barriuso, with David García-Selfa co-supervising the research internship at CESGA during which the initial research direction was proposed.

Additionally, this research project was made possible through the access granted by the Galicia Supercomputing Center (CESGA) to supercomputer FinisTerrae III and its permanent data storage system, which have been funded by the NextGeneration EU 2021 Recovery, Transformation and Resilience Plan, ICT2021-006904, and also from the Pluriregional Operational Programme of Spain 2014-2020 of the European Regional Development Fund (ERDF), ICTS2019-02-CESGA-3, and from the State Programme for the Promotion of Scientific and Technical Research of Excellence of the State Plan for Scientific and Technical Research and Innovation 2013-2016 State subprogramme for scientific and technical infrastructures and equipment of ERDF, CESG15-DE-3114.

## Declaration of Generative AI and AI-assisted Technologies in the Writing Process

During the preparation of this manuscript, the authors used Claude (Anthropic) to improve the language, grammar, readability, and clarity of argument presentation. After using this tool, the authors reviewed and edited the content as needed and take full responsibility for the content of the publication.

# References

[1] Ziming Liu, Yixuan Wang, Sachin Vaidya, Fabian Ruehle, James Halverson, Marin Soljačić, Thomas Y. Hou, and Max Tegmark. KAN: Kolmogorov-Arnold networks. In The Thirteenth International Conference on Learning Representations (ICLR), 2025. arXiv:2404.19756.

[2] A. N. Kolmogorov. On the representation of continuous functions of many variables by superposition of continuous functions of one variable and addition. Doklady Akademii Nauk, 114(5):953–956, 1957.

[3] V. I. Arnol'd. On functions of three variables. Doklady Akademii Nauk, 114(4):679–681, 1957.

[4] Ziming Liu, Pingchuan Ma, Yixuan Wang, Wojciech Matusik, and Max Tegmark. Kolmogorov-Arnold networks meet Science. Physical Review X, 15:041051, 2025. arXiv:2408.10205.

[5] Cale Coffman and Lizhong Chen. MatrixKAN: Parallelized Kolmogorov-Arnold Network. arXiv preprint arXiv:2502.07176, 2025.

[6] Andrew Polar and Michael Poluektov. Concurrent training methods for Kolmogorov-Arnold networks: Disjoint datasets and FPGA implementation. arXiv preprint arXiv:2512.18921v5, 2025.

[7] Priya Goyal, Piotr Dollár, Ross Girshick, Pieter Noordhuis, Lukasz Wesolowski, Aapo Kyrola, Andrew Tulloch, Yangqing Jia, and Kaiming He. Accurate, large minibatch SGD: Training ImageNet in 1 hour. arXiv preprint arXiv:1706.02677, 2017.

[8] Jianmin Chen, Xinghao Pan, Rajat Monga, Samy Bengio, and Rafal Jozefowicz. Revisiting distributed synchronous SGD. arXiv preprint arXiv:1604.00981, 2016.

[9] Dan Alistarh, Demjan Grubic, Jerry Li, Ryota Tomioka, and Milan Vojnovic. QSGD: Communication-efficient SGD via gradient quantization and encoding. In Advances in Neural Information Processing Systems, pages 1709–1720, 2017.

[10] Shen Li, Yanli Zhao, Rohan Varma, Omkar Salpekar, Pieter Noordhuis, Teng Li, Adam Paszke, Jeff Smith, Brian Vaughan, Pritam Damania, and Soumith Chintala. PyTorch distributed: Experiences on accelerating data parallel training. Proceedings of the VLDB Endowment, 13(12):3005–3018, 2020. arXiv:2006.15704.

[11] Pitch Patarasuk and Xin Yuan. Bandwidth optimal all-reduce algorithms for clusters of workstations. Journal of Parallel and Distributed Computing, 69(2):117–124, 2009.

[12] Mohammad Shoeybi, Mostofa Patwary, Raul Puri, Patrick LeGresley, Jared Casper, and Bryan Catanzaro. Megatron-LM: Training multi-billion parameter language models using model parallelism. arXiv preprint arXiv:1909.08053, 2019.

[13] Samyam Rajbhandari, Jeff Rasley, Olatunji Ruwase, and Yuxiong He. ZeRO: Memory optimizations toward training trillion parameter models. In SC20: International Conference for High Performance Computing, Networking, Storage and Analysis, pages 1–16. IEEE, 2020.

[14] Peter Mattson, Christine Cheng, Greg Diamos, Cody Coleman, Paulius Micikevicius, David Patterson, Hanlin Tang, Gu-Yeon Wei, Peter Bailis, Victor Bittorf, et al. MLPerf training benchmark. Proceedings of Machine Learning and Systems, 2:336–349, 2020.

[15] Baidu Research. DeepBench, 2016. Available at: https://github.com/baidu-research/DeepBench.

[16] Prajit Ramachandran, Barret Zoph, and Quoc V. Le. Searching for activation functions. arXiv preprint arXiv:1710.05941, 2017.

[17] Ameya D. Jagtap, Kenji Kawaguchi, and George Em Karniadakis. Adaptive activation functions accelerate convergence in deep and physics-informed neural networks. Journal of Computational Physics, 404:109136, 2020.

[18] Carl de Boor. A Practical Guide to Splines. Revised ed. Applied Mathematical Sciences, vol. 27. Springer, 2001. ISBN 978-0-387-95366-3.